\documentstyle[12pt,epsfig]{article}
\input epsf
\newcommand{\be}{\begin{eqnarray}}
\newcommand{\ee}{\end{eqnarray}}
\newcommand{\mpp}{M_{\pi\pi}}

\def\Dirac1#1{#1\hskip-2pt/}
\def\Dirac#1{#1\hskip-6pt/}

\def\beq{\begin{equation}}
\def\eeq{\end{equation}}
\begin{document}
%
%
\rightline{RUB-TP2-01/00}
\begin{center}
\begin{large}
{\bf Partonic structure of $\pi$ and $\rho$
mesons from data on hard exclusive
production of two pions off nucleon}\\[0.35cm]
\end{large}
{  B.~Clerbaux$^{a,*}$ and M.V. Polyakov$^{b,c}$}\\[0.3cm]
{\em $^a$ Inter-University Institute for High Energies  (U.L.B.), Brussels,
     Belgium } \\ [0.05cm]
{\em $*$ Since January 2000 at C.E.R.N., Geneva, Switzerland} \\ [0.05cm]
{ e-mail: Barbara.Clerbaux@cern.ch}\\[0.15cm]
{\em $^b$Petersburg Nuclear Physics Institute,
  188350, Gatchina, Russia}\\[0.05cm]
{\em $^c$ Institut f\"ur Theoretische Physik II, Ruhr-Universit\"at Bochum,
D-44780 Bochum, Germany} \\ [0.05cm]
{ e-mail: maximp@tp2.ruhr-uni-bochum.de}\\
\end{center}
\vspace{0.4cm}
%
\begin{abstract}
\noindent
We fitted the $\pi\pi$ mass distribution in the range
$0.5 \leq \mpp \leq 1.1$~GeV
measured in hard exclusive
positron-proton reactions at HERA by the form dictated by QCD at
leading twist level. Extracted parameters are related to valence quark
distribution in the pion, and
to the pion and $\rho$ meson distribution amplitudes.
We obtain, for the first time, a measurement of the second Gegenbauer
coefficient of the $\rho$ meson distribution amplitude:
$a_2^{(\rho)}= -0.10\pm 0.20 $ for a
photon virtuality of $\langle Q^2 \rangle=21.2$~GeV$^2$.
\end{abstract}

\noindent{ \bf 1. Introduction}\\[0.2cm]
Owing to QCD factorisation theorem for hard exclusive reactions
\cite{CFS}
the dependence of the amplitude  of the reaction
\footnote{With $Q^2\sim 2(p\cdot q)\gg \Lambda_{\rm QCD}^2$
 and $M_T^2,M_{T'}^2, (p'-p)^2, \mpp^2\ll Q^2. $}
\begin{equation}
\gamma^*_L(q)+T(p)\to \pi^+\pi^- + T'(p')
\label{proc2}
\end{equation}
with longitudinally polarised virtual photon
on the di-pion mass $\mpp$ factors out in an universal
(independent of the target) factor.
At leading order
in $\alpha_s(Q^2)$  this factor has the form
\cite{MVP98}:
\begin{equation}
{\cal A}(\mpp)\propto
\int_0^1 \frac{dz}{z}\,
\Phi^{I}(z,\zeta,\mpp;  Q^2)\;.
\label{f1}
\end{equation}
Here $\Phi^{I}(z,\zeta,\mpp;  Q^2)$ is the two-pion light cone
distribution amplitude ($2\pi$DA) \cite{DGPT},
which depends on
$z$--longitudinal momentum carried by the quark, $\zeta$
characterising the distribution of longitudinal momentum between
the two pions\footnote{For detailed definition
of kinematical variables $z$ and $\zeta$ see refs.~\cite{MVP98,DGPT}.},
and the invariant mass of produced pions $\mpp$,
the superscript $I$ standing for isospin of produced pions
($I=0,1$).
The dependence on the virtuality of the incident photon
$Q^2$ is governed by the ERBL evolution equation
\cite{ERBL}.

For the process (\ref{proc2}) at small $x_{Bj}=Q^2/2(p\cdot q)$
the production of two
pions in the isoscalar channel is strongly suppressed
relative to the isovector channel \cite{Ben}, because the former is
mediated by $C$-parity odd exchange.
At asymptotically large $Q^2$ QCD
predicts the following simple form for the isovector $2\pi$DA
\cite{MVP98,PW98}:
\begin{equation}
\Phi^{I=1}_{asym}(z,\zeta,\mpp)=
6  z(1-z) (2\zeta-1)F_\pi(\mpp)\; ,
\label{asym}
\end{equation}
where $F_\pi(\mpp)$ is the pion electro-magnetic (e.m.) form factor
in time-like region,
measured with high precision in low energy experiments \cite{barkov,cmd2}.
From eqs.~(\ref{f1},\ref{asym}) we conclude that at asymptotically large
$Q^2$ QCD predicts unambiguously the shape of the di-pion mass
distribution. Asymptotically the dependence of the amplitude
on $\mpp$ has the form:
\begin{equation}
{\cal A}_{asym}\propto
\beta(\mpp)\ F_{\pi}(\mpp)\ \cos\theta\; ,
\label{asyamp}
\end{equation}
where $\beta(\mpp)=\sqrt{1-\frac{4 m_\pi^2}{\mpp^2}}$ is
the velocity of pions in their centre of mass system (cms)
and $\theta$ is the angle between the directions of the positive pion
and the momentum of produced $\pi^+\pi^-$ system in the $\pi^+ \pi^-$ cms.
This angle is related to $\zeta$ in the following way:
\be
\cos\theta=\frac{2\zeta-1}{\beta(\mpp)}\; .
\ee
The corresponding
di-pion mass distribution has asymptotically the form
\begin{equation}
\frac{dN(\mpp)}{d\mpp}\propto
\mpp\ \beta(\mpp)^{3}
 |F_{\pi}(\mpp)|^2\, .
\label{asyshape}
\end{equation}
The asymptotic shape for any di-meson production (mesons $M_1$, $M_2$)
was derived in \cite{BFGMS,FKS,AFS}, where it was related to the
cross section of $e^+e^-\to M_1, M_2$ at low $\sqrt s$.

At non-asymptotic $Q^2$ values, the $2\pi$DA deviates from its asymptotic
form (\ref{asym}). This deviation can be described by a few parameters
which can be related to
quark distributions (skewed and usual) in the pion and to distribution
amplitudes of mesons ($\pi,\rho$, etc... ), for details see \cite{MVP98}.

\vspace{0.3cm}
\noindent{ \bf 2. Deviations from the asymptotic form }\\[0.2cm]
The first non-trivial deviation from the asymptotic form of the
$2\pi$DA occurs in $P$ and $F$ waves
\cite{MVP98}. Generically their effect on $\mpp$ dependence of the
hard amplitude can be written as:
\be
\nonumber
{\cal A}(\mpp)&\propto&
\beta(\mpp)\ e^{i\delta_1}|F_{\pi}(\mpp)|
\biggl(1 +D_1(\mpp) \biggr)
\ P_1(\cos\theta)\\ &+&
\beta(\mpp)^3\ e^{i\delta_3}D_{2}(\mpp)\ P_3(\cos\theta)
\; ,
\label{deviation}
\ee
where $P_l(\cos\theta)$ are Legendre polynomials and
$\delta_1(\mpp)$ and  $\delta_3(\mpp)$ are the $P$-wave and the
$F$-wave $\pi\pi$ scattering phase shifts,
which are well known from low-energy experiments.
The functions $D_{1,2}(\mpp)$ describe the deviation of the amplitude's $\mpp$
dependence from the asymptotic form.
These functions are real and can be parametrised as:
\begin{eqnarray}
\nonumber
D_1(\mpp,Q^2)&=&A_1(Q^2) e^{\bar b_1 \mpp^2}
-\frac{6 m_\pi^2}{\mpp^2}\
A_2(Q^2)
e^{\bar b_2 \mpp^2} \\
\nonumber
D_2(\mpp,Q^2)&=&A_2(Q^2) e^{b_3 \mpp^2}\, .
\end{eqnarray}
The dependence of $A_{1,2}(Q^2)$ on $Q^2$ is governed by the
QCD evolution and in leading order is given by:
\begin{eqnarray}
A_{1,2}(Q^2)=A_{1,2}(\mu_0)
\biggl(
\frac{\alpha_s(Q^2)}{\alpha_s(\mu_0)}
\biggr)^{50/(99-6 n_f)}\, .
\label{evolu}
\end{eqnarray}
With increasing $Q^2$, the parameters $A_{1,2}(Q^2)$ go
logarithmically to zero and one reproduces the asymptotic
formula (\ref{asyamp}).
The parameters $A_{1,2}(Q^2)$ are directly related to partonic structure
of $\pi$ and $\rho$ mesons, see section~4 and ref~\cite{MVP98}.

The parameter $b_3$ is $Q^2$-independent but is
$\mpp$ dependent. The latter dependence is fixed by $\pi\pi$
scattering phase shifts (see for derivation \cite{MVP98}):
\be
b_3(\mpp)= \bar b_3 + \mbox{\rm Re}
\frac{\mpp^2}{\pi}\int_{4m_\pi^2}^\infty
ds \frac{\delta_3(s)}{s^2(s-\mpp^2-i0)}\, .
\label{paramf}
\ee
In above equations $\bar b_i$ are
subtraction constants of corresponding dispersion relations for
functions $D_{1,2}(\mpp)$, see details in \cite{MVP98}.

Using expression (\ref{deviation}) we can derive the
 form of the $\mpp$ distribution:
\be
\nonumber
\frac{dN(\mpp)}{d\mpp}=N\Biggl[&&\beta(\mpp)^3\ \mpp
|F_{\pi}(\mpp)|^2 \biggl(1+D_1(\mpp)\biggr)^2+\\
\label{fg}
&&\frac 37 \ \mpp\ \beta(\mpp)^7\ D_2(\mpp)^2+ \\
&&{\rm higher\ waves\ } l\geq 5\
\nonumber
\Biggr] \ ,
\ee
where the higher partial waves can be safely neglected.

\vspace{0.3cm}
\noindent{ \bf 3. Angular distributions of produced pions}\\[0.2cm]
Another way to obtain sensitivity to the
non-asymptotic parameters $A_{1,2}(Q^2)$
is to study the
two pion angular distributions.
From the expression of the amplitude~(\ref{deviation}) one can derive
the $\mpp$ dependence of intensity densities
$\langle Y_l^m(\theta,\phi) \rangle$ defined as:
\be
\frac{d}{d\mpp}
\langle  Y_l^m   \rangle =
\mpp\ \beta(\mpp)\ \int_{-1}^1 d(\cos\theta )
\int_0^{2\pi} d\varphi
\  Y_l^m(\theta,\varphi)
|{\cal A}(\mpp,\theta,\varphi)|^2 \, ,
\ee
where $Y_l^m(\theta,\varphi)$ are spherical harmonics, $\varphi$
is the azimuthal angle between the pion decay plane and plane
formed by momentum of the virtual photon and total momentum
of produced $\pi^+\pi^-$ system.

For the first nontrivial intensity density
$\langle Y_2^0 \rangle$ we have:
\be
\nonumber
\frac{d}{d\mpp}
\langle Y_2^0   \rangle &\propto&
\mpp \Biggl[
\beta(\mpp)^3|F_{\pi}(\mpp)|^2\ \biggl(1+D_1(\mpp)\biggr)^2\\
\nonumber
&+& \frac{9}{7}\ \beta(\mpp)^5 \
|F_{\pi}(\mpp)|\ \biggl(1+D_1(\mpp)\biggr)\
D_2(\mpp)\
\cos\biggl[\delta_1(\mpp)-\delta_3(\mpp)\biggr]\\
\nonumber
&+& \frac{2}{7}\ \beta(\mpp)^7
D_2(\mpp)^2 \Biggr] \, .
\label{y20}
\ee
The combination $\langle Y_0^0-\sqrt 5/2 Y_2^0 \rangle $ is especially
sensitive to the deviations from asymptotic form because it is
exactly zero asymptotically:
\be
\frac{d}{d\mpp}
\nonumber
\langle  Y_0^0 -\frac{\sqrt 5}{2} Y_2^0 \rangle&\propto&
\mpp \Bigl[
  \beta(\mpp)^5 \ |F_{\pi}(\mpp)|\ \biggl(1+D_1(\mpp)\biggr)
\\
\nonumber
&\times& D_2(\mpp) \cos\biggl[\delta_1(\mpp)-\delta_3(\mpp)\biggr] \\
&-& \frac{1}{9}\ \beta(\mpp)^7
D_2(\mpp)^2 \Bigr] \, .
\label{y00y20}
\ee
Let us note, however that the expressions for intensity densities
obtained above are based on leading twist expression for the
amplitude~(\ref{deviation}), in particular, the contribution
of transversely polarised photon is neglected (it contributes
at higher twist level). Since in the combination of intensity densities
(\ref{y00y20}) the leading twist contribution tends to cancel,
the effect of higher twists (for which the cancellation is not expected)
can be numerically large.

The study of angular distribution
allows to make separation\footnote{Under assumption of
$s-$channel helicity conservation (SCHC),
which holds with good accuracy
experimentally \cite{h1_dis}.}
of productions by longitudinally
(leading twist) and
transversely (higher twist) polarised photons. Indeed, we can form
combinations of intensity densities which get contributions only from
$\gamma^*_L$, so we minimise the contributions of
higher twists.
One of such combinations
 \footnote{Strictly speaking
 this combination gets also
 contribution from transverse polarisation
 of the photon, but it occurs only in $F$-wave and therefore
 is expected to be small.}
is, for example,
\be
\frac{d}{d\mpp}
\nonumber
\langle  Y_0^0 +\sqrt 5\ Y_2^0 \rangle&\propto&
\mpp \Biggl[
\beta(\mpp)^3|F_{\pi}(\mpp)|^2\ \biggl(1+D_1(\mpp)\biggr)^2\\
\nonumber
&+& \frac{6}{7}\ \beta(\mpp)^5 \
|F_{\pi}(\mpp)|\ \biggl(1+D_1(\mpp)\biggr)\\
\label{y00y20p}
&\times&D_2(\mpp)\
\cos\biggl[\delta_1(\mpp)-\delta_3(\mpp)\biggr]\\
\nonumber
&+& \frac{1}{3}\ \beta(\mpp)^7
D_2(\mpp)^2 \Biggr] \, .
\ee
The details of the derivation and the
analysis of the data will be presented
elsewhere.

The advantage of the formalism presented in this section is that it
maximally uses the information on pion e.m. form factor $F_\pi(\mpp)$
and pion phase shifts $\delta_l(\mpp)$ at low $\mpp$ which are very
well known from low energy experiments. For example, from the known
phase shifts $\delta_1(\mpp)$ and $\delta_3(\mpp)$ we conclude
that the term proportional to
$\cos [\delta_1(\mpp)-\delta_3(\mpp) ]$ changes the sign
around $\mpp=0.8$~GeV, so to increase the sensitivity to this term
it would be interesting to consider a ``$\mpp$ asymmetry" for various
observables:
\be
\biggl(
\int_{2 m_\pi}^{0.8 } d\mpp - \int_{0.8}^{M_{\rm max} }
 d\mpp \biggr)
\frac{d}{d\mpp} ({\rm an\ observable})\, .
\ee

\vspace{0.3cm}
\noindent{ \bf 4. Expected values for the parameters}\\[0.2cm]
By crossing relations the parameter
$A_2$ is related to the third moment of the valence quark distribution
in the pion \cite{MVP98}:
\be
A_2(Q^2)=\frac 76 M_3(Q^2)= \frac 76
\int_0^1 dx\ x^2\ (u_{\pi^+} -\bar u_{\pi^+})\, .
\label{m3}
\ee
If the parametrisation suggested in~\cite{GL99} is used for the
quark distributions in the pion, we obtain the values of the parameter
$A_2(Q^2)$ listed in Table~\ref{table:num}.

\begin{table}[htb]
\begin{center}
\begin{tabular}{|c|c|c|}
\hline
$Q^2 ($GeV$^2$) & $A_2$ \\
\hline
2& 0.110   \\
4& 0.099   \\
10& 0.089  \\
15& 0.085  \\
\hline
\end{tabular}
\caption{{\it Values of the parameter $A_2$ as function of $Q^2$,
    obtained using eq.~(\ref{m3}) and quark distribution in the pion
    suggested in~\protect{\cite{GL99}}.}}
\label{table:num}
\end{center}
\end{table}

The value of $A_1$ is constrained by the soft pion theorem
\cite{MVP98}:
$$A_1(Q^2)=a_2^{(\pi)}(Q^2)-A_2(Q^2)\, , $$
where  $a_2^{(\pi)}$ is the second Gegenbauer coefficient of
the pion distribution amplitude.
Unfortunately this parameter is not very well measured.
In ref.~\cite{yak} the value of $a_2^{(\pi)}=0.19\pm 0.13$ at
$Q^2=2.4$~GeV$^2$ is quoted.

Additionally the parameters $A_1(Q^2)$ and $\bar b_1$
can be related to the second
Gegenbauer coefficient $a_2^{(\rho)}(Q^2)$
of the $\rho$ meson distribution amplitude as \cite{MVP98}:
\be
a_2^{(\rho)}(Q^2)=A_1(Q^2) e^{\bar b_1 \mpp^2}\,.
\label{a2rho}
\ee
Up to now there is no direct experimental information about
$a_2^{(\rho)}(Q^2)$.

The  $\mpp$ dependence of the parameter $b_3$ is estimated to be weak
in the range of $0.5<\mpp<1.1$~GeV and is then replaced by a constant
 $\bar b_3$.
The parameters $\bar b_i$ were estimated in the instanton model
of QCD vacuum~\cite{MVP98} to be around the following values:
\be
\nonumber
\bar b_1&=&-0.75\ {\rm \ GeV}^{-2}\\
\label{brange}
\bar b_2&=&-0.75\ {\rm \ GeV}^{-2}\\
\nonumber
\bar b_3&=&\ \ 0.75\ {\rm \ GeV}^{-2}.
\ee

\vspace{0.3cm}
\noindent{ \bf 5. Results of fits to the HERA data}\\[0.2cm]
Recently data on di-pion mass distribution in hard exclusive reaction
was measured at HERA by the reaction
\begin{equation}
e+p \rightarrow e+p+\pi^+ + \pi^-. \label{eq:reac}
\end{equation}
This section presents attempts to fit HERA data
by the leading twist parametrisations described in the sections above.

In order to study the $Q^2$ evolution of the di-pion mass distribution
three sets of data,
the ZEUS photoproduction data
($Q^2$ $\approx$ 0)~\cite{zeus_gp}, the ZEUS low $Q^2$ data
($0.25 < Q^2 < 0.85$ GeV$^2$)~\cite{zeus_dis} and the H1 high $Q^2$ data
($2.5 < Q^2 < 60$ GeV$^2$)~\cite{h1_dis} are used.
The mean W value for these samples is around 70 GeV, where $W$
is the energy in the
photon-proton cms ($W^2$ $\simeq$ $Q^2/x_{Bj} - Q^2$).
For the three samples, the di-pion mass distribution~\footnote{
  Note that the di-pion mass distributions measured at HERA
are not corrected for
the transverse photon production. Nevertheless for
$Q^2$ greater than 2 GeV$^2$ the longitudinal cross section dominates the
transverse cross section.}
was studied in the ranges $0.55 < \mpp < 1.2$ GeV,
$0.55 < \mpp < 1.1$ GeV and
$0.50 < \mpp < 1.1$ GeV, respectively. The main background
to eq.~(\ref{eq:reac})
consists of events in which the proton diffractively
dissociates into hadrons, and is estimated
to be around 20 \% for the ZEUS samples and around 10 \%  for the H1 sample.
This background
does not distort the di-pion mass distribution discussed in previous
sections if the mass of recoiled baryonic system is much smaller
than the hard scale $Q^2$.
The contamination from the production of
$\omega$ (decaying into $\pi^+ \pi^- \pi^0$)
and $\phi$ (decaying into $K^+ K^-$ or $\pi^+ \pi^- \pi^0$) mesons
was estimated to be few
percents for the ZEUS samples and 7 \% for the H1 sample.
This background is mainly situated at low mass
$\mpp <$ 0.6 GeV.

\begin{figure}[tp]
\begin{center}
\epsfig{file=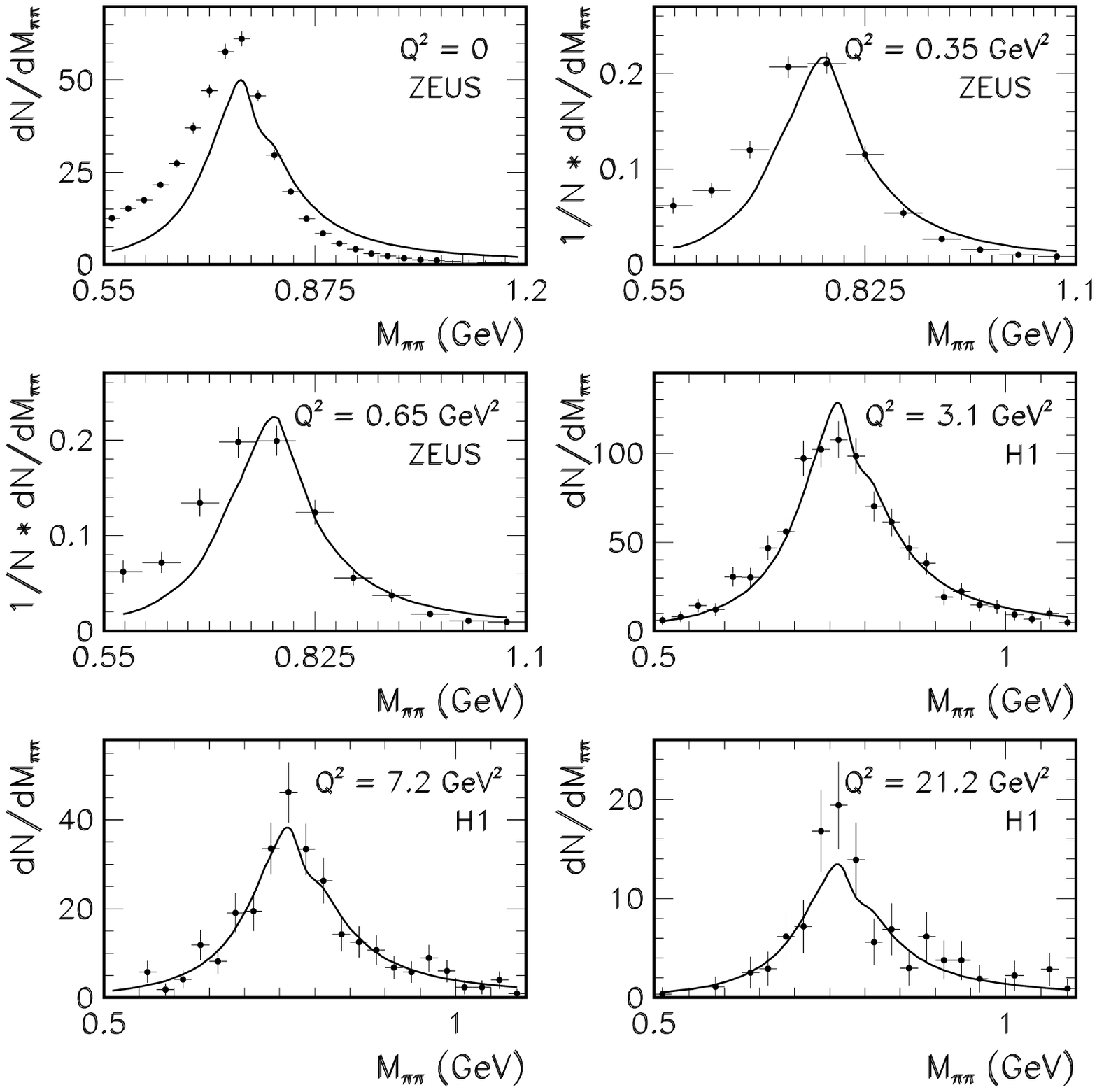,%
     height=14.0cm,width=14.0cm}
\caption{{\it Di-pion mass distribution for six $Q^2$ values.
     The black points correspond to measurements from the
     ZEUS~\protect\cite{zeus_gp,zeus_dis} and
     H1~\protect\cite{h1_dis} experiments. The curves show the
     result of the fit of the asymptotic
form~(\ref{asyshape}).}}
\label{fig:fit_1par}
\end{center}
\end{figure}

Fig.~\ref{fig:fit_1par} presents the di-pion mass distribution
(black points) for six $Q^2$ values:
the first distribution corresponds to the ZEUS photoproduction sample,
the two following ones to the
ZEUS low $Q^2$ sample and the three last ones to the H1 sample. For
the H1 data, the distributions are corrected bin per bin for the
production of $\omega$ and $\phi$ mesons.

Firstly, we attempt to fit the HERA data with the simplest formula of the
asymptotic form~(\ref{asyshape}) where the di-pion mass distribution is
directly proportional to the square of
the pion e.m. form factor. This latter was recently precisely measured
in low energy~\footnote{
  At leading twist the energy dependence factories out.
  The shape of the di-pion mass distribution is then energy independent in
  the present formalism.}
experiments~\cite{cmd2} (see also ~\cite{barkov}),
and we use directly these experimental data for $|F_{\pi}(\mpp)|$.
The result of the fits is superposed to the data in
Fig.~\ref{fig:fit_1par} (black curves).
The asymptotic form does not reproduce the distribution in photoproduction
and at low $Q^2$ as expected\footnote{The asymptotic
form~(\ref{asyshape}) is expected to work only in the hard regime
$Q^2 \gg \Lambda_{\rm QCD}^2$, also let us stress once more that the form
of di-pion mass distribution of eq.~(\ref{fg}) also makes sense
only in the hard regime.},
but describes well the data at high $Q^2$,
the $\chi^2/ndf$ of the fits for the six $Q^2$ bins
of Fig.~\ref{fig:fit_1par} being respectively
1625/25, 169/10, 69/10, 30.5/23, 16.6/21 and 11.6/18.
We see that di-pion mass distributions in soft (low $Q^2$) and
hard (large $Q^2$)  regimes are essentially different and are
described by different physics. Therefore the parametrisations
designed for soft processes \cite{Soding,Stod,NPW} are not
relevant for hard processes, and the values of the parameters
extracted from large $Q^2$ data are not related directly to
physical observables.

\begin{figure}[tp]
\begin{center}
\epsfig{file=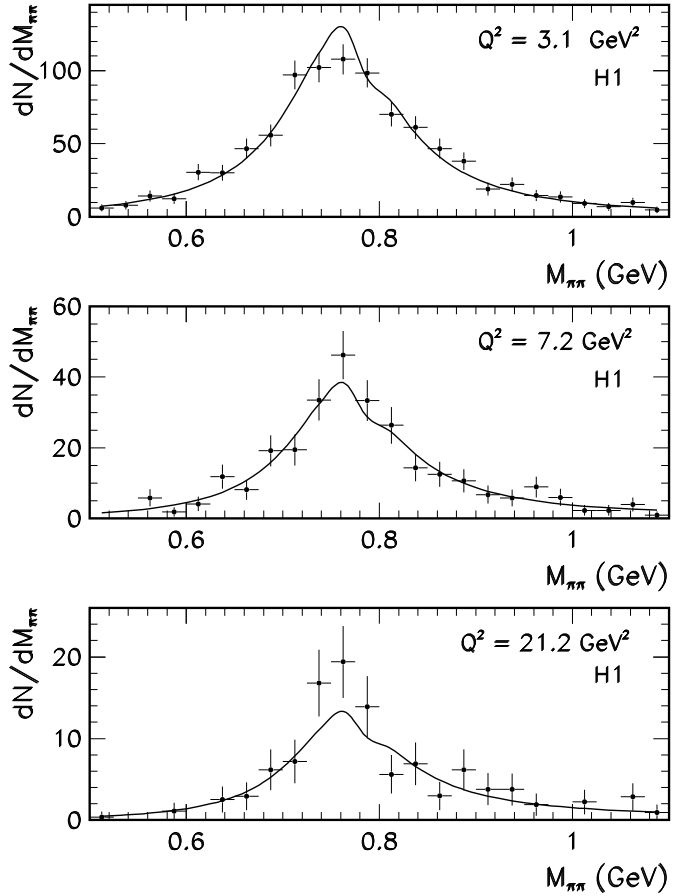,%
     height=14.0cm,width=14.0cm}
\caption{{\it Di-pion mass distribution for three $Q^2$ values.
     The black points correspond to measurements from the
     H1~\protect\cite{h1_dis} experiment and the curves show the
     result of the fit of eq.~(\ref{fg}).}}
\label{fig:fit_3par}
\end{center}
\end{figure}

Secondly, we study the deviations from the asymptotic form of the
di-pion mass distribution measured at HERA. As the formalism is valid
in the hard regime $Q^2\gg \Lambda_{\rm QCD}^2$,
only the H1 samples ($Q^2$
$>$ 2.5 GeV$^2$) are considered.  The parametrisation~(\ref{fg}) is
fitted to the data, with three free parameters: the normalisation $N$,
and the parameters $A_1$ and $A_2$.  For the parameters $\bar b_1, \bar
b_2$, and $\bar b_3$ we use the values given by the calculation in the
instanton model of QCD vacuum, see eq.~(\ref{brange}).  The result of
the fits is presented in Fig.~\ref{fig:fit_3par} as the black curves,
and the values obtained for the parameters are listed in
Table~\ref{table:param}. The three fits have a $\chi^2/ndf$ value
smaller than one.
\begin{table}[htb]
\begin{center}
\begin{tabular}{|l|l|l|}
\hline
  $Q^2$ = 3.1 GeV$^2$       &  $Q^2$ = 7.2 GeV$^2$       &  $Q^2$ = 21.2 GeV$^2$      \\
 \hline
  $N$   = 1.59 $\pm$ 0.21   &  $N$   = 1.11 $\pm$ 0.55   &  $N$   = 0.66 $\pm$ 0.30   \\
  $A_1$ = 1.10 $\pm$ 0.18   &  $A_1$ = 0.26 $\pm$ 0.49   &  $A_1$ = -0.16 $\pm$ 0.34  \\
  $A_2$ = -0.34 $\pm$ 0.50  &  $A_2$ = 0.22 $\pm$ 0.45   &  $A_2$ = 0.18 $\pm$ 0.41   \\
  $\chi^2/ndf$ = 20.6/21    &  $\chi^2/ndf$ = 16.5/19   &  $\chi^2/ndf$ = 11.3/16     \\
\hline
\end{tabular}
\caption{{\it Values of the normalisation $N$, the parameters $A_1$ and $A_2$,
     and the $\chi^2/ndf$ obtained from the fit of eq.~(\ref{fg}) to the H1
     data~\cite{h1_dis}, using the values given in eq.~(\ref{brange}) for the
     parameters $\bar b_1, \bar b_2$, and  $\bar b_3$.}}
\label{table:param}
\end{center}
\end{table}

We observe an indication for a decrease of the parameter
$A_1$ when $Q^2$ increases, what is expected from QCD evolution,
see eq.~(\ref{evolu}). In general the values of parameters $A_1$ and $A_2$
are in agreement with theoretical expectations (see section~4). However
the errors of the parameters $A_1$ and  $A_2$ are large, due to limited
statistic available.  Nevertheless the sensitivity of the data to
the parameter $A_1$ is encouraging. More precise data
from the 1997-99 years data taking at HERA would bring considerably more
accurate information on these parameters which are related directly
to the valence quark distribution in the
pion, and pion and $\rho$ meson distribution amplitudes.

\vspace{0.3cm}
\noindent{ \bf 6. Discussion of the results and conclusions}\\[0.2cm]
The analysis presented here is based on the
leading order, twist-2 formalism. The power corrections
(higher twist contributions) to the
amplitude of the reaction~(\ref{proc2}) are systematically
neglected. The size of higher twist corrections
might be rather large \cite{Bel2,VMS}, so that they constitute the
main theoretical systematic uncertainty in our analysis.
On general grounds the higher twist effects should be smaller
for such observables as the shape of the di-pion mass distribution
considered here, also the study of angular distribution of produced
pions as discussed in section~3 can be used to minimise the contributions
of higher twists.
Another source of theoretical errors is that we fixed values of
parameters $\bar b_i$ by the model values (\ref{brange}).
Hopefully more precise data
will allow us to measure these parameters.

To be on more safe side in respect to higher twist corrections
we use the highest values of
$\langle Q^2 \rangle=21.2$~GeV$^2$ available in
the data sample to evaluate
observables related to partonic structure of the pion and $\rho$ meson.
The results for the third moment of valence quark distribution in the
pion $M_3$ (see eq.~(\ref{m3})), the second Gegenbauer
coefficients of the pion and $\rho$ meson distribution amplitudes ($a_2^{(\pi)}$ and
$a_2^{(\rho)}$ respectively) at $\langle Q^2 \rangle=21.2$~GeV$^2$ are
presented in Table~\ref{table:result}.

\begin{table}[htb]
\begin{center}
\begin{tabular}{|l|l|l|}
\hline
  Quantity       & Our analysis       &  Other sources
\\
 \hline
  $M_3$          &  0.15 $\pm$ 0.35   &  0.07 \cite{GL99},\
$0.085\pm 0.005$ \cite{SMRS} \\
  $a_2^{(\pi)}$  &  0.02 $\pm$ 0.50
&  0.14$\pm$ 0.09 \cite{yak}  \\
  $a_2^{(\rho)}$ &  -0.10$\pm$ 0.20   &  ---  \\
\hline
\end{tabular}
\caption{{\it Values of the third moment of valence quark distribution
     in the pion $M_3$, and the second Gegenbauer coefficient of the
     pion and $\rho$ meson distribution amplitudes ($a_2^{(\pi)}$,
     $a_2^{(\rho)}$) at $\langle Q^2 \rangle=21.2$~GeV $^2$ obtained in
     our analysis, and in~\protect\cite{GL99,yak,SMRS}}.}
\label{table:result}
\end{center}
\end{table}

Although, due to the limited statistic, the error bars for
physical observables are large we see that the obtained values are
in agreement with other experiments.
This agreement is especially interesting because
previously the restrictions for $a_2^{\pi}$ and $M_3$ were obtained
from completely different measurements: the value of second Gegenbauer
coefficient $a_2^{\pi}$ was obtained in ref.~\cite{yak} from analysis of
data on $\gamma \pi$ transition form factor at large $Q^2$, whereas
the third moment $M_3$ is restricted by $\pi N$ Drell-Yan
data~\cite{GL99,SMRS}\footnote{Lattice calculations of $M_3$
are in agreement with results of analysis of \cite{GL99,SMRS}:
$0.09\pm 0.03$ \cite{Beck} and $0.10\pm 0.02$ \cite{MaSa}
at $\langle Q^2 \rangle=21.2$}.
We see that the formalism
to extract the partonic structure of the pion and $\rho$ mesons
suggested here is complementary to already known methods.

Our analysis allowed us to obtain
for the first time an
experimental estimate of the second Gegenbauer coefficient of
the $\rho$ meson distribution amplitude
$a_2^{(\rho)}=-0.1\pm 0.2$ at $\langle Q^2\rangle=21.2$~GeV$^2$.
Unfortunately the precision of determination of $a_2^{(\rho)}$
is still low to discriminate between different model predictions
for this quantity  (QCD sum rule:
$0.12\pm 0.06$ \cite{BaBr}, $0.05\pm 0.01$ \cite{BaMi} and
instanton model: $-0.07$ \cite{MVP98,PW99}).
With increasing of statistics the accuracy of
determination of the parameters related to the partonic structure of
the mesons can be considerably improved. Also the studies of
angular distributions of produced pions as discussed in section~3
can considerably increase the sensitivity to parameters of partonic
structure of pions and their resonances.

As a final remark we note that the formalism developed here can
be easily generalised
to the case of production of other mesons
in hard exclusive reactions
(e.g. $K\bar K$, $3\pi, 4\pi, K\bar K \pi\pi$, etc.). The study of
meson mass and angular distributions of produced meson can provide us
with rich information on partonic structure of these mesons and their
resonances. Interesting prediction can be made \cite{BFGMS,FKS,AFS}
for the asymptotic shape
of meson cluster mass distribution for the reaction:
\be
\gamma^*_L + p \to (hadrons) +N'\, ,
\ee
where the mass of the hadron cluster $M_X^2\ll Q^2$. In this case at
asymptotically large $Q^2$ we have:

\be
\label{rr}
\frac{dN}{d M_X^2}&\propto& R(M_X), \ \ \ {\rm with}\\
\nonumber
R(\sqrt s)&=&\frac{\sigma(e^+e^- \to hadrons)}{\sigma(e^+e^- \to \mu^+\mu^-)}\,.
\ee
It would be interesting to check experimentally this asymptotic formula
and try to detect deviations from it. Generically, this deviation
is described by the vacuum correlator of two light-ray opertors.
We showed here that the asymtotic
expression (\ref{rr}) works pretty well at low $M_X$ (for $Q^2 > 7$~GeV$^2$)
where the $2\pi$ channel dominates over all other hadronic channels,
it would be interesting to check how good this formula works at
higher $M_X$. Also it would be interesting to see whether the formula
like (\ref{rr}) is applied also for multimeson partial cross section.

\vspace{0.5cm}
\noindent
{\bf Acknowledgements:}
\noindent
We gratefully acknowledge discussions with
K. Goeke, B.~Lehmann-Dronke,
L.~Mankiewicz, G. Piller,  A.~Sch\"afer, A.G.~Shuvaev,
W. Weise, and C.~Weiss. Special thanks are due to Lyonya Frankfurt and
Mark Strikman for suggestion to look into the problem and many
valuable contributions.
The work of B.C. is supported by
the National Foundation for Scientific Reseach (FNRS).
M.V.P. is supported by RFBR grant 96-15-96764, by DFG and BMBF.
%

\newpage
\newpage

\end{document}